\documentstyle[12pt]{article}
  \advance\oddsidemargin  by -1in
  \advance\evensidemargin by -1in
  \advance\topmargin      by -0.5in
\def\lsim{\mathrel{\rlap{\lower4pt\hbox{\hskip1pt$\sim$}}
    \raise1pt\hbox{$<$}}}         
\def\gsim{\mathrel{\rlap{\lower4pt\hbox{\hskip1pt$\sim$}}
    \raise1pt\hbox{$>$}}}         

\def\beq{\begin{equation}}
\def\endeq{\end{equation}}
\def\arr{\begin{eqnarray}}
\def\endarr{\end{eqnarray}}
\makeindex

\begin{document}
\phantom{.}\hspace{8.5cm}{\bf KFA-IKP(Th)-1993-27}\\
\phantom{.}\hspace{9cm}{\sl 8 November 1993}
\vspace{2.5cm}\\
\begin{center}
{\Huge \bf  Decisive test of
color transparency in exclusive electroproduction of
vector mesons \vspace{1.5cm}\\ }
{\Large B.Z.Kopeliovich$^{1)}$, J.Nemchick$^{1,2)}$,
N.N.Nikolaev$^{3,4}$\\
and B.G.Zakharov$^{3}$\bigskip\\ }
{\sl
$^{1)}$Laboratory of Nuclear Problems, JINR, Dubna, Russia\\
$^{2)}$Institute of Experimental Physics, Slovak Academy of
Sciences, \\Watsonova 47, 043 53 Kosice, Slovak Republik   \\
$^{3)}$L.D.Landau Institute for Theoretical Physics, GSP-1, 117940,\\
ul.Kosygina 2, V-334, Moscow, Russia\\
$^{4)}$IKP(Theorie), KFA J\"ulich, D-52425 J\"ulich, Germany
\vspace{1.0cm}\\}

{\bf \Large A b s t r a c t}
\bigskip\\
\end{center}

The exclusive production of vector mesons in deep inelastic
scattering is a hard scattering process with the well controlled
size of quark configurations which dominate the production
amplitude. This allows an unambiguous prediction of
color transparency effects in the coherent and incoherent
production of vector mesons on nuclei.
We demonstrate how the very mechanism of color transparency
leads to a belated onset of color transparency effects as
a function of $Q^{2}$.
We conclude that the
$Q^{2}$ dependence of the exclusive $\rho^{0}$-meson production
on nuclei and nucleons observed in the Fermilab E665 experiment
gives a solid evidence for the onset of color transparency.
We propose the scaling relation between the $\rho^{0}$ and
the $J/\Psi$ production, which further tests the mechanism of color
transparency in exclusive (virtual) photoproduction.
 \vspace{1.5cm}\\
\centerline{\sl Submitted to Physics Letters B}
\bigskip\\
\centerline{\bf E-mail: kph154@zam001.zam.kfa-juelich.de}
\pagebreak



The exclusive production of vector mesons $\gamma^{*}p\rightarrow Vp$
in deep inelastic scattering
is a hard scattering process in which the transverse
size $r_{Q}$ of quark configurations which dominate the production
amplitude is under the good control [1] ,
\beq
r_{Q} = {2 \over \sqrt{m_{V}^{2}+Q^{2}}} \, ,
\label{eq:1}
\endeq
which makes this reaction an ideal laboratory for testing
color transparency (CT) ideas (for the recent review on
CT see [2,3]). In the exclusive production off nuclei,
the nuclear attenuation
effects were predicted [1] to be $\propto r_{Q}^{2}$ and vanish at
large virtuality $Q^{2}$ of the photon, which was confirmed by
the recent data on the $\rho^{0}$ production from the
FNAL E665 experiment [4]. In [1] we have presented
quantitative predictions
for the incoherent production of $\rho^{0}$-mesons  on
iron nucleus. Since the E665 data
give the first definitive signal of CT in the
hard scattering process,
an analysis of the further consequences
of CT for virtual photoproduction of vector
mesons is called upon. In this paper we present
predictions for the signal of CT in the
coherent production on nuclei and extend calculations [1]
for the incoherent production to wider range of nuclei,
using the formalism developed by us earlier [1,5,6].
We demonstrate how the very mechanism of CT
leads to a belated onset of CT effects as
a function of $Q^{2}$, in good agreement with the
E665 data [4]. We propose the scaling law which relates CT
effects in the (virtual) photoproduction of the $\rho^{0}$
and the $J/\Psi$ mesons.

We begin with a brief outline of the lightcone approach to
exclusive $\gamma^{*}N\rightarrow VN$ production developed in [5,6,1,2].
At the high energy $\nu$ of the virtual photon
the reaction mechanism
greatly simplifies: The photon fluctuates into the $q\bar{q}$ pair at
a large distance (the coherence length)
\begin{equation}
l_{c}={{2\nu} \over Q^{2}+m_{V}^{2}} \, .
\label{eq:2}
\end{equation}
in front of the target nucleon (nucleus). After interaction  the
$q\bar{q}$ pair recombines into the vector meson $V$ with the
recombination (formation) length
\begin{equation}
l_{f}={\nu \over m_{V}\Delta m} \, ,
\label{eq:3}
\end{equation}
where $\Delta m $ is the typical level splitting in the
quarkonium. At high energy $\nu$ both $l_{c}$ and $l_{f}$  greatly
exceed the radius $R_{N}$ ($R_{A}$) of the target nucleon (nucleus),
and the transverse
size $\vec{r}$ of the $q\bar{q}$ pair and the longitudinal
momentum partition $z$ and $(1-z)$ between the quark and antiquark
of the pair do not change during the interaction with the target.
This enables one to introduce the lightcone wave function
$\Psi_{\gamma^{*}}(\rho,z)$ of the $q\bar{q}$ fluctuation
[7]. The color-singlet  $q\bar{q}$ pair
interacts with the target nucleon with the cross
section [7,8]
\beq
\sigma(r) = {\pi^{2}\over 3}r^{2}
{\cal F}(\nu,r)  \, ,
\label{eq:4}
\endeq
where ${\cal F}(\nu,r)$ is related at small $r$ to the gluon
structure function of the proton, ${\cal F}(\nu,r)=
\alpha_{S}(r)xg(x,Q_{r}^{2})$, evaluated at
$Q_{r}^{2} \sim    1/r^{2}$ and the value of the Bjorken variable
$x\approx (Q_{r}^{2}+m_{V}^{2})/
2m_{N}\nu$, and
$\alpha_{S}(r)$ is the running QCD coupling. This factor
${\cal F}(\nu,r)$
takes into account the effect of higher $q\bar{q}g_{1}...g_{n}$
Fock components in the lightcone wave function
of the photon and the vector meson [8].
It is a smooth function of $r$ compared to $r^{2}$ in
Eq.~(\ref{eq:4}). The cross section $\sigma(r)$
has the CT property of vanishing at
$r\rightarrow 0$  [9,10]. Since the CT property is
conveniently quantified in terms of $\sigma(r)$ (for the
review see [2,3]), it is
important to understand how $\sigma(r)$ is probed in the
exclusive electroproduction.

In order to set up the reference frame notice [1,2,5,6], that
the amplitude of the forward photoproduction
$\gamma^{*}N\rightarrow VN$ can be cast
in the quantum-mechanical form [$\vec{q}$ is the momentum transfer]

\beq
M(VN,\vec{q}=0)=\langle V|\sigma(r)|\gamma^{*}\rangle
=\int_{0}^{1} dz \int d^{2}\vec{r}\sigma(r)\Psi_{V}(r,z)^{*}
\Psi_{\gamma^{*}}(r,z)   \, .
\label{eq:5}
\endeq
The most important feature of
$\Psi_{\gamma^{*}}(r,z)$ is an exponential decrease
at large distances [7],
\beq
\Psi_{\gamma^{*}}(r,z)
\propto \exp(-\varepsilon r)  \, ,
\label{eq:6}
\endeq
where
\beq
\varepsilon^{2} = m_{q}^{2}+z(1-z)Q^{2} \approx
{1\over 4}(m_{V}^{2} + 4z(1-z)Q^{2})   \, .
\label{eq:7}
\endeq
In the nonrelativistic quarkonium $z \approx 1/2$,
$\Psi_{\gamma^{*}}(r,z)$ is concentarted at $r \lsim r_{Q}$ and
the wave function (6) [7] and
Eq.~(\ref{eq:1}) for $r_{Q}$ [1] fulfill the dream [11]
of having the
well specified control of the size of quark configurations
important in the hard scattering process.

The generalization to nuclear targets is straightforward [1,5,6].
In the incoherent production $\gamma^{*}A\rightarrow VA^{*}$ one
sums over all excitations of the final nucleus.
The nuclear transparency  for the incoherent production
equals [12,2,5,6]
\beq
T_{A}={\sigma_{A} \over A\sigma_{p}}={1\over A}
\int d^{2}\vec{b} T(b)
{\langle V |\sigma(r)
\exp\left[-{1\over 2} \sigma(r)T(b)\right] |\gamma^{*}
\rangle^{2} \over
\langle V|\sigma(r)|\gamma^{*}\rangle^{2} } =
1-\Sigma_{V} {1\over A}\int d^{2}\vec{b}T(b)^{2} +...
\label{eq:8}
\endeq
where $T(b)=\int dz n_{A}(b,z)$ is the optical thickness
of a nucleus at the impact parameter b,
the nuclear density $n_{A}(b,z)$ is
normalized to the nuclear mass number A,
$\int d^{3}\vec{r} n_{A}(\vec{r}) = A$,(for the compilation of
the nuclear density parameterizations see [13]). The observable
[2]
\beq
\Sigma_{V}={\langle V|\sigma(r)^{2}|\gamma^{*}\rangle
\over \langle V|\sigma(r)|\gamma^{*}\rangle } \,.
\label{eq:9}
\end{equation}
measures the strength of intranuclear final state interaction (FSI).

The amplitude of the coherent nuclear production
$\gamma^{*}A \rightarrow VA$ equals [9,12]
\arr
M(VA,\vec{q}) =
2\int d^{2}\vec{b}
\langle V |1-
\exp\left[-{1\over 2} \sigma(r)T(b)\right] |\gamma^{*}
\rangle \exp(-i\vec{q}\,\vec{b})   \nonumber\\
= AM(VN,\vec{q}=0) [G_{em}(q)-
\Sigma_{V} {G_{2}(q)\over 2A}\int d^{2}\vec{b}T(b)^{2} +... ]
\label{eq:10}
\endarr
Here $G_{em}(q)$ is the charge form factor of the nucleus and
the form factor of double scattering $G_{2}(q)$ is defined by
$\int d^{2}\vec{b}T(b)^{2}\exp(-i\vec{q}\,\vec{b})=
G_{2}(q)\int d^{2}\vec{b}T(b)^{2}$.
The total
cross section of the coherent production equals
\arr
\sigma_{coh}(VA) =
4\int d^{2}\vec{b}
\left|\langle V |1-
\exp\left[-{1\over 2} \sigma(r)T(b)\right] |\gamma^{*}
\rangle\right|^{2}\nonumber\\ =
\left|M(VN,\vec{q}=0)\right|^{2}\int d^{2}\vec{b}
T(b)^{2}\left[1-
{1\over 2}\Sigma_{V}T(b) +... \right]
\label{eq:11}
\endarr
For the sake of clarity in the subsequent discussion of the
onset of CT, in Eqs.~(\ref{eq:8},\ref{eq:10},\ref{eq:11}) we
have explicitly shown the leading term of FSI.
The signature of the coherent production
is a sharp forward diffraction peak with the
slope $B_{coh} \approx {1\over 3}R_{ch}(A)^{2} \gg B(VN)$,
see Eq.~(\ref{eq:10}), whereas in
the incoherent
nuclear production the $\vec{q}\,^{2}$-dependence is to a good
accuracy the same as in production on the free nucleons
([14], for the review
see [15]).
This allows to separate the coherent and incoherent cross
sections even at limited resolution in $\vec{q}\,^{2}$.

Now we turn to implications of Eq.~(\ref{eq:1}) and of
CT (\ref{eq:4}) for the production
rate and nuclear FSI effects.
The wave function of the vector meson is smooth at small $r$.
Because of CT property (\ref{eq:4}) the  integrand
of (\ref{eq:5}) is $\propto r^{3}\exp(-{r\over r_{Q}})$ and
the amplitude of (virtual)
photoproduction will be dominated by contribution
from size
\beq
r_{S} \sim 3r_{Q}  \, ,
\label{eq:12}
\endeq
which falls in into the perturbative QCD domain $r_{S} \ll R_{V}$
at a sufficiently large $Q^{2}$. In this domain for production of the
transversely polarized vector mesons by the transversely polarized
photons $\gamma_{T}^{*} N \rightarrow V_{T}N$
we have an
estimate
\beq
M_{T}(VN,\vec{q}=0)
\propto {r_{S}^{2} \over R_{V}^{3/2}}\sigma(r_{S})
\propto {1\over (Q^{2}+m_{V}^{2})^{2}}
{\cal F}(\nu,r_{S}) \, .
\label{eq:13}
\endeq
The $(m_{V}^{2}+Q^{2})^{-2}$ behavior of the amplitude (\ref{eq:13})
is different from the VDM prediction [15,16]. This difference
comes from the factor $\sigma(r_{S})\sim    1/(Q^{2}+m_{V}^{2})$
in (\ref{eq:13}) which
emphasizes a relevance of CT property of
$\sigma(r)$ to the total production rate.

The evaluation of the strength of FSI $\Sigma_{V}$ goes as follows:
The integrand of the matrix element
$\langle V|\sigma(r)^{2}|\gamma^{*}\rangle$ is
$\sim r^{5}\exp(-{r\over r_{Q}})$ and
is peaked at
\beq
r\sim r_{FSI}=(4-5)r_{Q} \, .
\label{eq:14}
\endeq
(Extension to the higher-order rescatterings is
straightforward.)
This gives an estimate
\beq
\Sigma_{V} \approx \sigma(r_{FSI}) \,\, .
\label{eq:15}
\endeq
CT and/or weak FSI set in when $\Sigma_{V} \ll
\sigma_{tot}(VN) \approx \sigma(R_{V})$, {\sl i.e.,} when
$r_{FSI} \ll R_{V}$.
 Remarkably,
the large numerical factor $\approx (4-5)$ in the  r.h.s.
of Eq.~(\ref{eq:14}) comes from CT property of $\sigma(r)$,
and the very CT property of the production mechanism
predicts a belated onset of the CT effect in
nuclear attenuation, which
requires $r_{Q} \ll {1\over 5}R_{V}$. Notice, that in the
opposite to the
absolute production rate (see Eq.~(13)), in the regime of CT
$\Sigma_{V}$ is insensitive to
the wave function of the vector meson. Thus,
predictions of attenuation effects are much less model dependent,
and in this paper we concentrate on the nuclear transparency.
(At a small and moderate $Q^{2}$, when $r_{FSI}\sim R_{V}$, the
observable $\Sigma_{V'}$ for the production of the radially
excited vector mesons $V'$
is extremely sensitive to the nodal
structure of the wave function of the $V'$,
which may lead $\Sigma_{V'} <0$ and
to the antishadowing phenomenon [1,5,6].)
We predict that
$1-T_{A}$ scales with $r_{Q}^{2}$ [1], i.e., with
$(Q^{2}+m_{V}^{2})$ :
\beq
1-T_{A} \propto
{A\over R_{ch}(A)^{2}}r_{FSI}^{2} \sim
{A\over R_{ch}(A)^{2}}{1\over Q^{2}+m_{V}^{2}}  \, ,
\label{eq:16}
\endeq
where $R_{ch}(A)$ is the charge radius of the nucleus.
The simple law (16) holds at $1-T_{A} \ll 1$.

Before the comparison of the free-nucleon and the nuclear production,
we emphasize that the mechanism of the free-nucleon reaction changes
drastically from $Q^{2}\lsim m_{V}^{2}$ to $Q^{2}\gg m_{V}^{2}$.
As Sakurai, Fraas and Shildknecht [16] have emphasized
long ago, from the essentially kinematical considerations one
finds a dominance of the longitudinal cross section
at large $Q^{2}$,
\beq
{\sigma_{L} \over \sigma_{T}} \approx {Q^{2} \over m_{V}^{2}} \, ,
\label{eq:17}
\endeq
which agrees with the recent data from the E665 [4]
and NMC [17] collaborations. Here we just mention that the
relationship (\ref{eq:17}) holds for the nonrelativistic
quarkonium, and the relativistic effects slow down the rise
(\ref{eq:17}) [18].

The relativistic effects also slow down the
rapid decrease (\ref{eq:13})
of $M_{T}$ with $Q^{2}$ [18]. Specifically, because of
the CT property of $\sigma(r)$ large values of $r$ are favored
in the integrand of the production amplitude. The photon
wavefunction $\Psi_{\gamma^{*}}(r,z)$ admits large $r$ if
$\varepsilon$ Eq.~(\ref{eq:7}) is small, {\sl i.e.,} if either
$z$ or $(1-z)$ is small. Such an asymmetric quark pairs in the
vector meson have a large
intrinsic longitudinal momentum, which suppresses the
wavefunction of the vector meson.
Nonetheless, at very large $Q^{2}$
the production amplitude will be dominated by the asymmetric
quark configurations in the vector meson, rather than by the
nonrelativistic configurations $z\sim 1/2$.
Even with allowance for the relativistic corrections,
FSI effects in $\sigma_{L}$
and $\sigma_{T}$ differ little. We predict the
approximately $A$-independent
polarization
density matrix of the produced $\rho^{0}$ mesons.
The E665 data [4] confirm this.
The detailed
discussion of the relativistic effects in the
$\rho^{0}$ production goes beyond the scope of the present paper
and will be presented elsewhere [18]. Here we just emphasize
that because $\Psi_{\gamma^{*}}(r,z)$ is well understood [7],
the virtual
photoproduction offers a unique opportunity of scanning the
wave function of vector mesons [1].

Regarding the experimental definition of the exclusive production,
the word of caution is in order. The coherent production selects
the truly exclusive production and its unambiguous signature is a
very narrow diffraction peak.
In the opposite to that, because of the limited energy resolution and
imperfect rejection of the particle production
in the target vertex, the
incoherent sample is contaminated by inelastic interactions
$\gamma^{*}p\rightarrow Vp^{*}$ with excitation of the
target proton. (Such an inelastic background was a major
problem in earlier
measurements of the exclusive $\rho^{0}$ production in deep
inelastic scattering [19].) There are good reasons to
expect that the ratio of the
inelastic background to the exclusive reaction on the free and
bound nucleons will be approximately the same. Also, in the regime
of CT the cross section $\sigma(r)$ has a universal dependence
on $r$, which is insensitive to the transition in the target nucleon
vetrex (for instance, see the considerations in [7,8]). Consequently,
the inelastic background should only
weakly affect the $A$- and $Q^{2}$-dependence of nuclear
trasnparency for the incoherent production. A
comparison of coherent cross sections on two
nuclei is ambiguity free. However, the relative normalization of the
coherent nuclear cross section to the hydrogen cross section depends
on the inelastic background in the latter. For this reason we
compare $\sigma_{coh}(A)$ and $\sigma_{coh}(C)$.

In Fig.1 we show our predictions for
the nuclear transparency for the incoherent
cross section as a function of $Q^{2}$. (Predictions for production
on the iron nucleus were published [1] before the E665 data were
reported.)
. Nuclear attenuation
is very strong at small $Q^{2}$ and gradually decreases with
$Q^{2}$. This rise of nuclear transparency $T_{A}$ with $Q^{2}$
is particularly dramatic for the heavy
nuclei ($Ca,\, Pb$), and leaves no doubts that the onset of CT
is observed.  Notice, that even at the highest
$Q^{2} \sim 10$ GeV$^{2}$ of the E665 experiment we predict
rather strong attenuation effect: although at this value
of $Q^{2}$ Eq.~(\ref{eq:1}) gives $r_{Q} \approx 0.13$ f ,
the Eq.~(\ref{eq:14}) shows that FSI is dominated by quark
configuration having the relatively large transverse size
$r_{FSI} \sim 0.5$ f. The large value of $r_{FSI}$ also
shows that the relativistic effects are not yet important
in the nuclear attenuation calculations.
Our predictions for the $Q^{2}$
dependence of the nuclear transparency are in good
agreement with the E665 data [4]. These data were taken at
$\nu \approx 200$ GeV, so that
the frozen size condition
$l_{c},l_{f} >R_{A}$ is fulfilled over the whole kinematical range
of the E665 experiment.

In Fig.~2 we present our predictions for the $Q^{2}$ dependence
of nuclear transparency for the forward coherent production on nuclei
\beq
T_{A}^{(coh)} = \left. {d\sigma_{A}^{(coh)} \over A^{2} d\sigma_{N}}
\right|_{\vec{q}\,^{2}=0}\, .
\label{eq:18}
\endeq
We predict a rise of $T_{A}^{(coh)}$ with $Q^{2}$. Even at
$Q^{2}\sim 10$ GeV$^{2}$ we predict a substantial departure
from $T_{A}^{(coh)}=1$ expected for the complete CT.
The experimental
determination of the absolute value of the forward production
cross section is difficult as it
requires high $\vec{q}\,^{2}$-resolution. Because of the
above mentioned problem of the inelastic background in the
hydrogen data we present predictions for the $A/C$ ratio.

The total coherent
production cross section is much less sensitive to the
$\vec{q}\,^{2}$-resolution.
In Fig.~3 we present our predictions for the $Q^{2}$ dependence
of the coherent production cross section relative
to the cross section for the carbon nucleus.
For the regime of complete CT and/or vanishing FSI,
\beq
R_{coh}^{(CT)}(A/C)= {12 \sigma_{A} \over A\sigma_{C}}
\approx {AR_{ch}(C)^{2} \over 12R_{ch}(A)^{2}}\, ,
\label{eq:19}
\endeq
which gives $R_{coh}^{(CT)}(Ca/C)= 1.56$ and
$R_{coh}^{(CT)}(Pb/C) = 3.25$.
We find a good quantitative agreement
with the $Q^{2}$ dependence observed in the E665 experiment.
As we have emphasized above, the coherent production is
free of the inelastic background, and the E665 data
on the $Pb/C$ ratio give a particularly unambiguous evidence for
the onset of CT.

The (approximate) $A^{\alpha}$ parametrization is a convenient
short-hand representation of the $A$-dependence of nuclear
cross sections. The so defined exponent $\alpha$ slightly
depends on
the range of the mass number $A$ used in the fit.
Then, Eq.~(13) predicts that $\alpha_{inc}(Q^{2})$
and $\alpha_{coh}(\vec{q}\,^{2}=0)$
tend to 1 and 2
from below, as $Q^{2}$ increases. In the limit of vanishing
final state interaction Eqs.~(15,16,18) predicts
$\sigma_{coh} \propto A^{2}/R_{ch}(A)^{2} \sim    A^{4/3}$, so
that $\alpha_{coh}(Q^{2})$ tends to $\approx {4\over 3}$
from below as $Q^{2}$ increases.
In Fig.4 we compare our estimate for the exponent
$\alpha$
with the results
of the $E665$ fits.
(Our exponent $\alpha$ is defined as an average of the two
values found from the ratio of theoretical prediction for the
$Pb/C$  and $Ca/C$ cross section ratios; the uncertainties in
$\alpha_{inc}$, $\alpha_{coh}(\vec{q}\,^{2}=0)$ and $\alpha_{coh}$
can be estimated as $\pm 0.03, \, \pm 0.03$ and
$\pm 0.05$, respectively. The A-dependence of the no-FSI coherent
cross section in the $C-Pb$ range of nuclei corresponds to
the exponent $\alpha_{coh}
 \approx 1.39$ at $Q^{2}\rightarrow \infty$
)
Both the $\alpha_{coh}(Q^{2})$ and
$\alpha_{inc}(Q^{2})$ rise with $Q^{2}$, which is still another
way of stating that the E665 data confirm the onset of CT.

The scaling law (\ref{eq:16}) predicts that
at $Q^{2} \approx 9$ GeV$^{2}$
the nuclear attenuation
for the incoherent $\rho^{0}$ production must be the same
as for the incoherent real ($Q^{2}=0$)
photoproduction of the $J/\Psi$.
The available $\rho^{0}$ and $J/\Psi$ production
data were taken with somewhat different nuclear targets.
For the $\rho^{0}$ production
at $\langle Q^{2} \rangle =7$ GeV$^{2}$ the
E665 experiment gives $[T_{Pb}/T_{C}]_{\rho^{0}} = 0.6 \pm 0.25$.
This can be
compared with the NMC result $[T_{Sn}/T_{C}]_{J/\Psi}= 0.7 \pm 0.1$ for
the real photoproduction of the $J/\Psi$ in the similar energy range.
The substantial departure of the above values of $T_{A}/T_{C}$
from unity confirms our result (\ref{eq:14},\ref{eq:15})
that even at $Q^{2}\sim 10$ GeV$^{2}$
the FSI is controlled by a large $r_{FSI} \sim 0.5$ f and is not
yet vanishing.

Similar scaling relationship holds for the coherent production of
the $\rho^{0}$ and the $J/\Psi$.
In the regime of complete CT
Eqs.~(\ref{eq:10},\ref{eq:11}) give [6]
$R_{coh}(Sn/C)=2.76$, $R_{coh}(Fe/Be)=2.82$, $R_{coh}(Pb/Be)=
4.79$. The experimental data on the real photoproduction
of $J/\Psi$ give a solid evidence for nonvanishing FSI:
$R_{coh}(Sn/C)=2.15\pm 0.10$ in the NMC experiment [21] and
$R_{coh}(Fe/Be)=2.28\pm 0.32$, $R_{coh}(Pb/Be)=3.47 \pm 0.50$
in the Fermilab E691 experiment [22]. In all cases the
$\approx 25\%$
departure
of the observed ratios for the $J/\Psi$ from predictions for the
complete CT
is of the same magnitude as in the highest $Q^{2}$
bin of the E665 data on the $\rho^{0}$ production (Fig.2).
The analysis [5,6] within the same formalism as used in
the present paper, gave  a good quantitative description
of the above $J/\Psi$ production data.
Higher precision data on the $J/\Psi$ and $\rho^{0}$ production
at higher $Q^{2}$ would be very interesting for further
tests of our scaling
law (\ref{eq:16}).

Our prediction (\ref{eq:13}) of a rapid decrease with $Q^{2}$
of the
total production cross section which also follows from CT,
is consistent with the experiment [17,19].
At very at large $Q^{2}$ the
production cross section and the ratio $\sigma_{L}/\sigma_{T}$
do gradually become sensitive to the relativistic components
of the wave function of vector mesons,
but in view of Eq.~(12) we predict a slow
onset of the relativistic effects.
As a matter of fact, the cross section (\ref{eq:4}) is well
understood theoretically (for the discussion of how the small-$r$
behavior of
$\sigma(r)$ is probed in deep inelastic scattering at
small $x$ see [7,8]). Henceforth, one rather must use
Eq.~(5) for the $Q^{2}$-controlled scanning [1] and measurement
of the wave function of vector mesons and for testing our
understanding of these wave functions in the relativistic
domain. To this end, the E665 results can be looked at as an
important probe of $\sigma(r)$ from $r\sim R_{V}$ down to
$r=r_{FSI}\sim 0.5{\rm f}$ at $Q^{2}=7 \,{\rm Gev}^{2}$.
The observed attenuation corresponds
to $\sigma(r=0.5{\rm f})\sim 8{\rm mb}$, and the
rise of the nuclear attenuation
towards small $Q^{2}$ and $r\sim R_{V}$ proves that
$\sigma(r)$ starts saturating at $\sigma(r) > \sigma_{tot}(\pi N)$
only at $r \gsim 1 {\rm f}$, in
agreement with the scenario [7].

We conclude that the solid signal of color transparency is seen
in the exclusive $\rho^{0}$ production in deep inelastic
scattering [4].
This is the fisrt quantitative confirmation of color transparency
ideas [9-11] in the hard scattering process (for the early
evidence for CT in the $\pi^{-}A\rightarrow \pi^{0}A^{*}$ charge
exchange reaction see [23]).
The E665 data are in good agreement with
predictions of the lightcone approach to the virtual photoproduction
of vector mesons developed by us [1,2,5,6]. Quite nontrivial
feature of our approach is that the very mechanism of CT enforces
rather large transverse size $r\sim r_{FSI}$
of the quark configurations which participate the intranuclear
FSI when travelling through the nucleus.
The scaling law (\ref{eq:16}) enables one to relate
CT effects in the $\rho^{0}$ and the $J/\Psi$ production.
The both $\rho^{0}$ and $J/\Psi$ sets of the experimental data
show that even at $Q^{2} \sim 10$ GeV$^{2}$ the residual FSI
is not yet vanishing, and confirm our prediction
of the slow onset of CT.
\bigskip\\
{\bf Acknowledgements:} One of the authors (N.N.N.) benefited
much from discussions on the E665 data with G.Fang and H.Schelmann.
We are grateful to C.Mariotti and A.Sandacz for
communications on the NMC data. The work of B.G.Zakharov was
supported in part by the grant from the Soros Foundation and
the DFG grant. Thanks are due to J.Speth for the hospitality
extended to N.N.N. and B.G.Z. at the Instutit f\"ur Kernphysik,
KFA, J\"ulich.

\pagebreak

\pagebreak
{\bf \LARGE Figure captions:}

\begin{itemize}

\item[Fig.~1 - ]
Predictions of nuclear transparency $T_{A}=\sigma_{A}/A\sigma_{N}$
for the incoherent
exclusive production of $\rho^{0}$ mesons vs. the E665 data [4].

\item[Fig.~2 - ]
Predictions of nuclear transparency
$T_{A}^{(coh)}/T_{C}^{(coh)}
=[144d\sigma_{A}/A^{2}d\sigma_{C}]_{\vec{q}\,^{2}=0}$
for the forward coherent production of the $\rho^{0}$ mesons.

\item[Fig.~3 - ]
Predictions of the $Q^{2}$ dependence of the
ratio of cross sections
$R_{coh}(A/C)=
12\sigma_{A}/A\sigma_{C}$
for coherent production of the $\rho^{0}$ mesons vs. the E665
data [4]. The arrows indicate predictions for the complete CT
at $Q^{2}\rightarrow \infty$.

\item[Fig.~4 - ]
Predictions of the $Q^{2}$ dependence of exponents of
parametrizations \\$\sigma_{A}(inc) \propto A^{\alpha_{inc}}$,~~
$[d\sigma_{A}(coh)/dq^{2}]_{\vec{q}\,^{2}=0} \propto
A^{\alpha_{coh}(\vec{q}\,^{2}=0)}$ and $\sigma_{A}(coh) \propto
A^{\alpha_{coh}}$
vs. the E665
data [4].

\end{itemize}
\end{document}